\documentclass[11pt,twoside]{article} 
\usepackage{cspm-asp2006}
\usepackage{epsfig,graphicx,natbib,url} 
\usepackage{lscape} 
\begin{document}
\setcounter{page}{337}

\markboth{Labrosse et al.}{Active Prominences}
\title{Spectral Diagnostics of Active Prominences}
\author{Nicolas Labrosse$^1$, Pierre Gouttebroze$^2$ and Jean-Claude Vial$^2$}
\affil{$^1$Institute of Mathematical and Physical Sciences, Aberystwyth, UK\\
       $^2$Institut d'Astrophysique Spatiale, Orsay, FR}

\begin{abstract}
  Active prominences exhibit plasma motions, resulting in difficulties
  with the interpretation of spectroscopic observations. These solar
  features being strongly influenced by the radiation coming from the
  solar disk, Doppler dimming or brightening effects may arise,
  depending on which lines are observed and on the velocity of the
  plasma. Interlocking between the different atomic energy levels and
  non local thermodynamic equilibrium lead to non-trivial spectral
  line profiles, and this calls for complex numerical modeling of the
  radiative transfer in order to understand the observations. We
  present such a tool, which solves the radiative transfer and
  statistical equilibrium for H, \ion{He}{i}, \ion{He}{ii}, and
  \ion{Ca}{ii} in moving prominences where radial plasma motions are
  taking place. It is found that for isothermal, isobaric prominence
  models, the \ion{He}{ii} resonance lines are very sensitive to the
  Doppler effect and thus show a strong Doppler dimming. The
  \ion{Ca}{ii} lines 
  Doppler effect for the prominence models considered here.  We
  illustrate how the code makes it possible to retrieve the plasma
  thermodynamic parameters by comparing computed and observed line
  profiles of hydrogen and helium resonance lines in a quiescent
  prominence.  This new non-LTE radiative transfer code including
  velocities allows us to better understand the formation of several
  lines of importance in prominences, and in conjunction with
  observations, infer the prominence plasma thermodynamic properties
  and full velocity vector.
\end{abstract}

\section{Introduction}

The main motivation of our modeling work is to contribute to building
realistic prominence models. For this, we need an accurate knowledge
of thermodynamic quantities (temperature, densities, \ldots), level
populations (useful, e.g., to infer the magnetic field properties from
spectro-polarimetric observations), velocity fields, energy
budget. However these quantities still have large uncertainties
associated with them. Observations of several different lines from
different atoms/ions allow us in theory to measure these
parameters. Among these lines, the H and He lines are important as
they are strong and largely contribute to the radiative
losses. However the prominence plasma being out of LTE and optically
thick in H and He resonance lines, the interpretation of line spectra
or intensities in radially moving prominences is a non-trivial
task. Therefore, non-LTE radiative transfer calculations including
velocity fields are needed to build realistic prominence models. Here
we present such calculations and preliminary results.

\section{Modeling Procedure}

The prominence is represented by a 1D plane-parallel slab standing
vertically above the solar surface. Each prominence model is defined
by a set of free parameters: the temperature $T$, the gas pressure
$p$, the slab thickness $L$ (or the total column mass), the height of
the slab above the limb $H$, the microturbulent velocity, and the
radial speed. For this preliminary study we consider isothermal and
isobaric prominences, although the code allows for inclusion of a
transition region between the cold prominence and the hot corona.  We
first solve the pressure equilibrium, the ionization equilibrium, and
the coupled statistical equilibrium (SE) and radiative transfer (RT)
equations for a 20 levels H atom. Then the SE and RT equations are
solved for other elements: \ion{He}{i} (29 levels) and \ion{He}{ii} (4
levels), and \ion{Ca}{ii} (5 levels).  More details on the modeling
of the hydrogen, calcium, and helium spectra in quiescent prominences
can be found in \citet{nl-gl00,nl-gh02,nl-lg04} respectively, and
references therein.

For the modeling of active and eruptive prominences, we use a
velocity-dependent incident radiation as boundary conditions for the
RT equations. It has already been shown by \citet{nl-hr87} in the case
of the hydrogen lines that the Doppler effect induces a frequency
shift of the incident profile relative to the rest case, and also a
distortion of the incident profile due to the variation of the Doppler
shift with the direction of the incident radiation. It is also the
case for the helium (\cite{nl-iau06}) and calcium incident radiation.

\section{Diagnostics of Radial Velocities}

\subsection{Integrated intensities}

We reproduce the results of \citet{nl-gvg97b,nl-gvg97a} who computed
the hydrogen radiation emitted by a radially moving prominence, using
partial redistribution in frequency (PRD) for the Lyman lines. We
obtain the same variation of the relative intensities (intensities
normalised to the line intensities when the prominence is at rest) and
the same line profiles for Ly$\alpha$, Ly$\beta$, and H$\alpha$. The
main result is that there exists an important coupling between
Ly$\beta$ and H$\alpha$ which causes these lines to be first Doppler
brightened, and then Doppler dimmed, with increasing velocity, while
there is only a Doppler dimming effect on Ly$\alpha$.

\begin{figure}
  \centering
  \plottwo{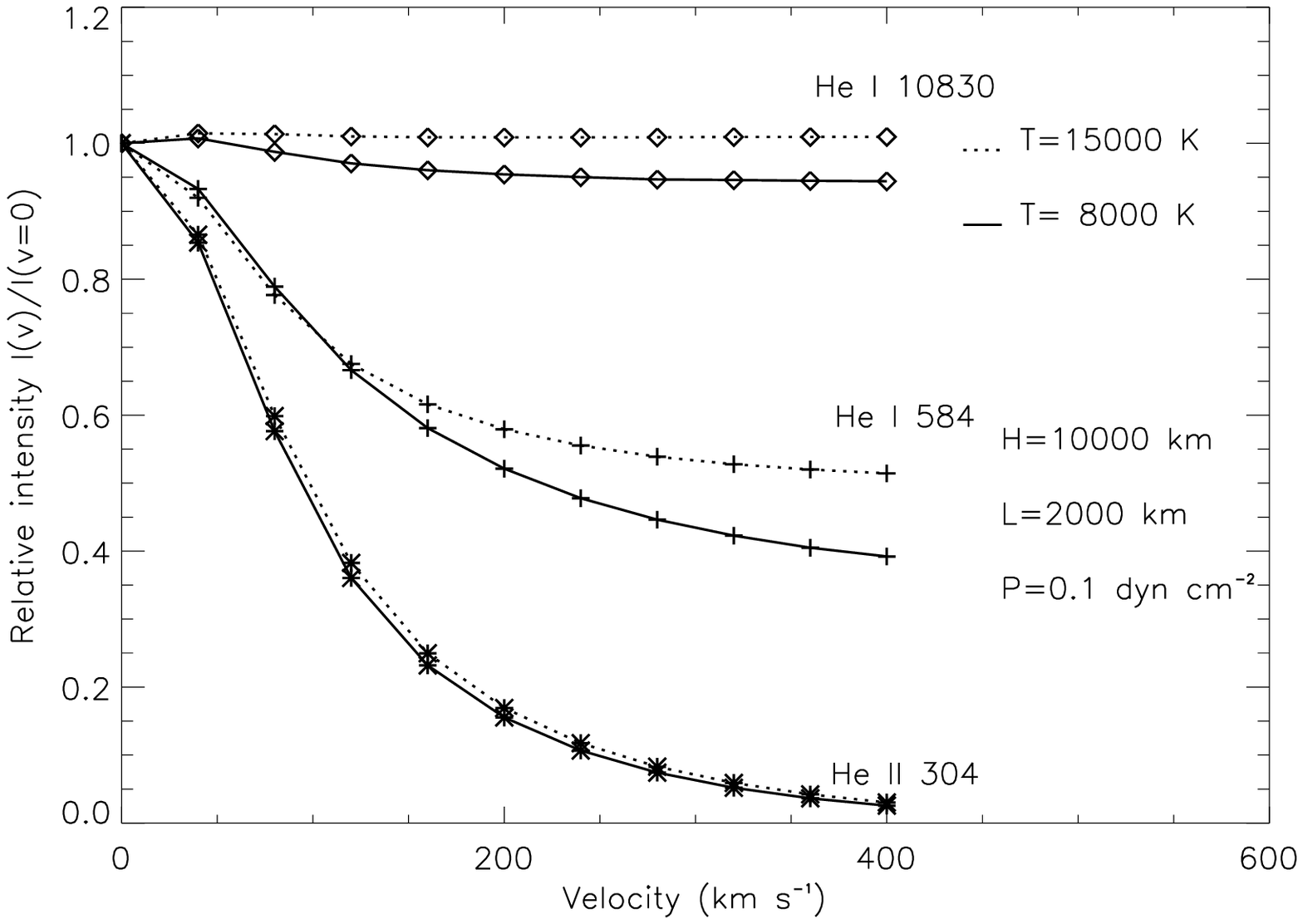}{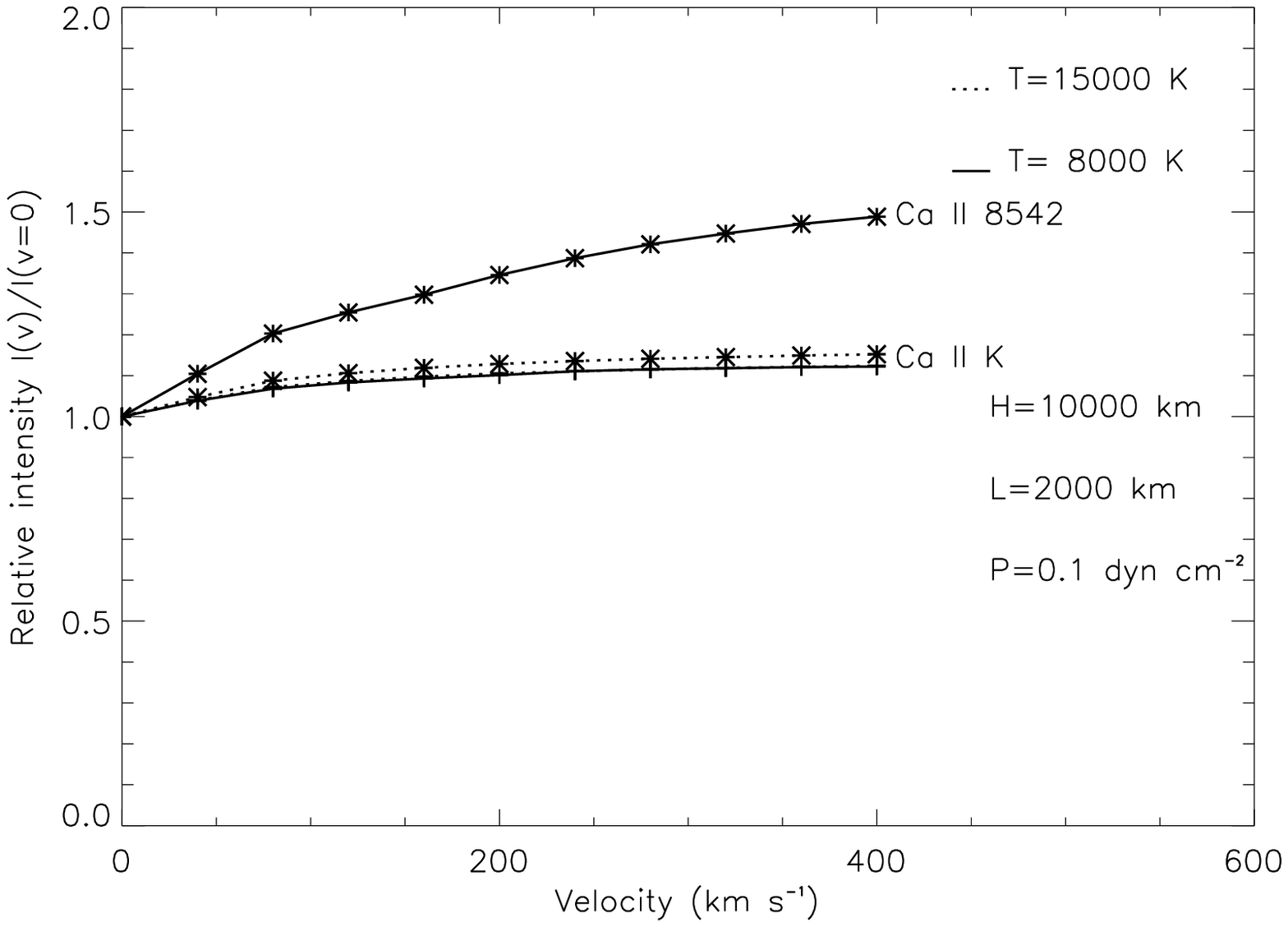}
  \caption[]{Relative intensities (intensities normalized to the line
  intensity at rest) as a function of velocity at 8000 K (solid lines)
  and 15000 K (dotted lines) for left: \ion{He}{i} 10830\,\AA\
  (diamonds), \ion{He}{i} 584\,\AA\ (plus), and \ion{He}{ii} 304\,\AA\
  (stars); right: \ion{Ca}{ii} K (plus) and \ion{Ca}{ii} 8542\,\AA\
  (stars).}
  \label{nl-fig:intensities}
\end{figure}

Figure~\ref{nl-fig:intensities} presents relative intensities as a
function of velocity for the \ion{He}{i} 584\,\AA, \ion{He}{ii}
304\,\AA, and \ion{He}{i} 10830\,\AA\ lines (left panel) and
\ion{Ca}{ii} K and \ion{Ca}{ii} 8542\,\AA\ lines (right panel) at two
different temperatures (8000\,K and 15000\,K).

The \ion{He}{i} 10830\,\AA\ line does not show any sensitivity to the
Doppler effect, which is mainly due to the very weak incident
absorption line. The \ion{He}{i} 584\,\AA\ line is quite sensitive to
the Doppler effect. Its Doppler dimming is more important at low
temperature. The \ion{He}{ii} resonance lines are the most sensitive
to the radial velocity of the plasma (the relative intensity of the
\ion{He}{ii} 256\,\AA\ line, not shown, exhibits a similar variation as
\ion{He}{ii} 304\,\AA), and the Doppler dimming is strong at the
temperatures considered in this study.  Such a result was expected
since the main mechanism of formation at these temperatures for these
lines is the scattering of the incident radiation
(\cite{nl-lg01}). Let us stress that in this preliminary study we have
not included a transition region between the cold prominence and the
hot corona (PCTR). The presence of a hotter plasma in the PCTR may
somehow decrease the sensitivity of the \ion{He}{ii} resonance lines
to the Doppler effect as collisions will become more important in the
formation processes of these lines. This will be investigated in a
future work.

The right panel of Fig.~\ref{nl-fig:intensities} indicates that there
is no strong Doppler effect on the \ion{Ca}{ii} resonance lines, while
we observe some Doppler brightening of the 8542\,\AA\ line (and indeed
of the other two infrared lines at 8498 and 8662\,\AA, not shown) at
low temperature.

\subsection{Line profiles}

\begin{figure}
  \centering
  \includegraphics[width=0.69\textwidth]{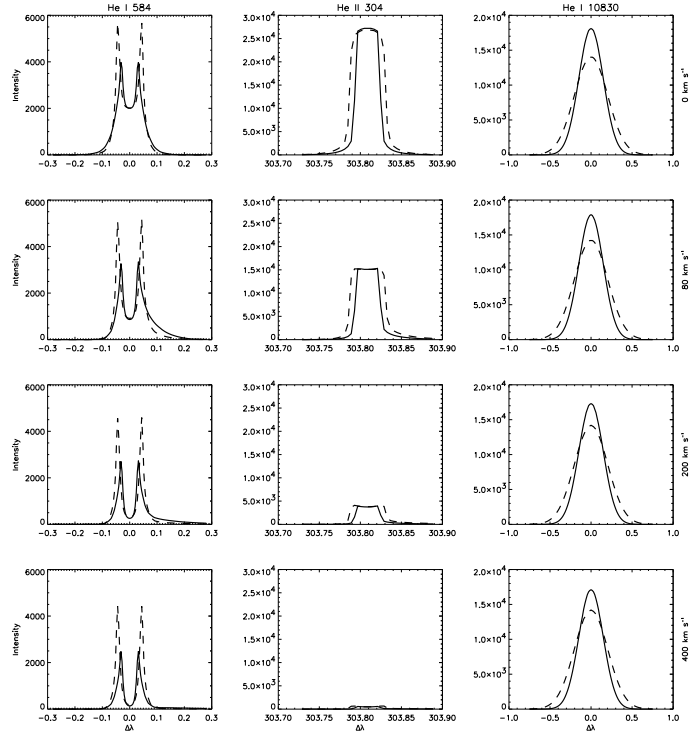}
  \caption[]{Line profiles for $T=8000$\,K (solid line) and $T=15\,000$\,K
  (dashed line), with $p=0.1$~dyn\,cm$^{-2}$, and $L=2000$~km, at
  different velocities: 0, 80, 200, and 400\,km\,s$^{-1}$ from top to
  bottom. Abscissa is $\Delta\lambda$ in\,\AA\ and vertical axis is
  specific intensity in
  erg\,s$^{-1}$\,cm$^{-2}$\,sr$^{-1}$\,\AA$^{-1}$. From left to right:
  \ion{He}{i} 584\,\AA, \ion{He}{ii} 304\,\AA, and \ion{He}{i}
  10830\,\AA.}
  \label{nl-fig:profils he}
\end{figure}

If spectroscopic observations of erupting prominences are available,
then a comparison between computed and observed line profiles can be
made. We show in Figs.~\ref{nl-fig:profils he} and \ref{nl-fig:profils
ca} the line profiles for the same helium and calcium lines considered
in Fig.~\ref{nl-fig:intensities} at two different temperatures (solid
line: 8000 K, dashed line: 15000 K), at four different velocities
(from top to bottom: 0, 80, 200, and 400\,km\,s$^{-1}$).

The Doppler dimming effect is well observed in the helium resonance
lines at 584\,\AA\ and 304\,\AA\ as the radial velocity is increased
(Fig.~\ref{nl-fig:profils he}). We can observe asymmetries in the line
profiles of these lines when the prominence plasma is moving radially,
with some intensity enhancement which is especially visible in the red
wing of the \ion{He}{i} 584\,\AA\ line at low temperature. This is
explained as follows. The radiation emitted by the disk center in our
code is represented by symmetrical line profiles. When the prominence
is at rest, the incident radiation illuminating the prominence slab is
symmetrical; however when the prominence plasma is moving radially the
incident line profile becomes distorted and shifted towards the
red. As we used the PRD standard approximation in our calculations of
the resonance lines of helium, the resulting line profiles are
asymmetrical. This would not have been the case if we had considered
complete redistribution in frequency (CRD) instead of PRD. The line
asymmetry is more visible for the \ion{He}{i} 584\,\AA\ line as its wings
are fairly bright, while the wing intensities of the \ion{He}{ii} 304\,\AA\
line are low. Despite the fact that the line asymmetry increases with
speed for both lines, it is more visible at low speeds (when the
intensity in the wing is high enough).  Finally, it is more pronounced
at low temperatures because the scattering of the incident radiation
is relatively more important as compared to collisional processes than
it is at higher temperatures.

\begin{figure}
  \centering
  \includegraphics[width=0.69\textwidth]{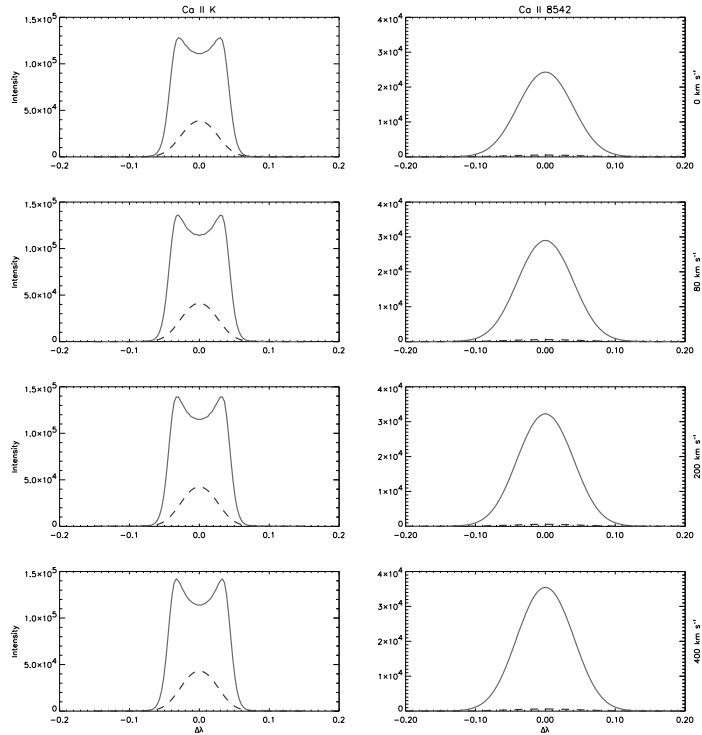}
  \caption[]{Same as in Fig.~\ref{nl-fig:profils he} for \ion{Ca}{ii} K and \ion{Ca}{ii} 8542\,\AA\ lines.}
  \label{nl-fig:profils ca}
\end{figure}

Figure~\ref{nl-fig:profils ca} shows that the intensities of the
\ion{Ca}{ii} lines are lower at 15\,000\,K than at 8000\,K, an effect of
\ion{Ca}{ii} to \ion{Ca}{iii} ionization (\cite{nl-gh02}).

\section{Diagnostics of Thermodynamic Parameters}

\begin{figure}
  \centering
  \includegraphics[width=0.54\textwidth]{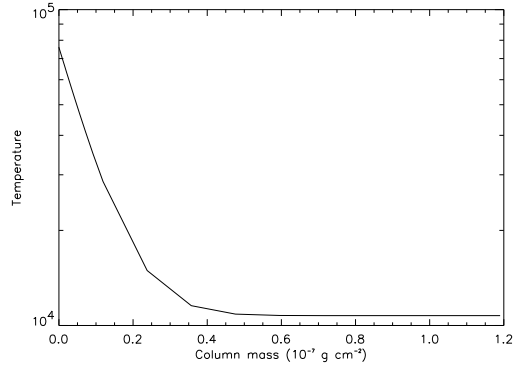}
  \caption[]{Temperature as a function of the column mass in one half of the prominence slab.}
  \label{nl-fig:temp}
\end{figure}

We show here how the non-LTE radiative transfer calculations can help
us to infer the thermodynamic properties of a prominence observed by
the SUMER spectrometer on SOHO. This prominence was actually a rather
quiet prominence and we have not included any velocity fields in these
calculations.  It was observed during the 13th MEDOC campaign held at
IAS in June 2004.  We select a few pixels in the SUMER slit which cut
across the prominence and average the line profiles there. We consider
the line profiles of two H resonance lines (Ly$\beta$ and
Ly$\epsilon$) and the \ion{He}{i} resonance line at 584\,\AA.  For the
comparison between computed and observed line profiles we now include
the presence of a PCTR.  The temperature variation between the cold
prominence core and the corona suggested by \citet{nl-ah99} has been
adopted for this study. By a trial and error process we selected the
temperature profile shown in Fig.~\ref{nl-fig:temp}. The other
prominence parameters are $p=0.023$~dyn\,cm$^{-2}$, $L=194$~km (total
column mass 2.4\,$10^{-7}$\,g\,cm$^{-2}$), $H=113722$~km, and the
microturbulent velocity $\xi=18$~km\,s$^{-1}$.

\begin{figure}
  \centering
  \includegraphics[width=0.69\textwidth]{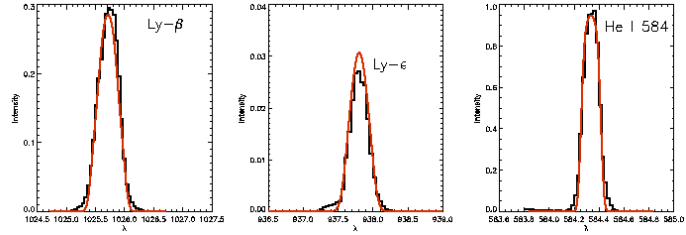}
  \caption[]{Observed (histograms) and computed (curves) lines profiles for Ly$\beta$ (left), Ly$\epsilon$ (middle), and \ion{He}{i} 584\,\AA\ (right).}
  \label{nl-fig:match}
\end{figure}

We obtain a very good agreement between the computed profiles
(convoluted with the SUMER instrumental profile) and the observed
profiles, as shown in Fig.~\ref{nl-fig:match}. It is worth noting that
fitting hydrogen \textit{and} helium resonance lines simultaneously
places strong constraints on the parameter space, and it was not
possible to find another set of parameters for the prominence that
would be significantly different than what is given above and that
would lead to a satisfactory fit of the observed profiles.

\section{Conclusions}

The non-LTE radiative transfer modeling that we are developing is a
key tool for interpreting observations and constructing realistic
prominence models.  The combination of lines from hydrogen, helium,
and calcium, places strong constraints on the models.  Imaging and
spectroscopy must be used for comparisons with calculations to
determine thermodynamic parameters and velocities.  The radial
velocity determined from the comparison between observed and computed
line profiles, in combination with line-of-sight velocities, should
allow us to infer the full velocity vector of the prominence plasma.
In a future work we will compare our model results with simultaneous
observations of, e.g., H$\alpha$ and \ion{He}{ii}~304\,\AA.

\acknowledgements N.\,L. acknowledges financial support from the
organisers of the Coimbra Solar Physics Meeting, the University of
Wales through the Gooding Fund, and PPARC through grant
PPA/G/O/2003/00017.


\begin{thebibliography}{}

\bibitem[\protect\astroncite{{Anzer} \& {Heinzel}}{1999}]{nl-ah99}
{Anzer} U., {Heinzel} P., 1999,
  \aap~  349, 974

\bibitem[\protect\astroncite{{Gontikakis} et~al.}{1997a}]{nl-gvg97b}
{Gontikakis} C., {Vial} J.-C., {Gouttebroze} P., 1997a,
  \aap~  325, 803

\bibitem[\protect\astroncite{{Gontikakis} et~al.}{1997b}]{nl-gvg97a}
{Gontikakis} C., {Vial} J.-C., {Gouttebroze} P., 1997b,
  \solphys 172, 189

\bibitem[\protect\astroncite{{Gouttebroze} \& {Heinzel}}{2002}]{nl-gh02}
{Gouttebroze} P., {Heinzel} P., 2002,
  \aap~  385, 273

\bibitem[\protect\astroncite{{Gouttebroze} \& {Labrosse}}{2000}]{nl-gl00}
{Gouttebroze} P., {Labrosse} N., 2000,
  \solphys 196, 349

\bibitem[\protect\astroncite{{Heinzel} \& {Rompolt}}{1987}]{nl-hr87}
{Heinzel} P., {Rompolt} B., 1987,
  \solphys 110, 171

\bibitem[\protect\astroncite{{Labrosse} \& {Gouttebroze}}{2001}]{nl-lg01}
{Labrosse} N., {Gouttebroze} P., 2001,
  \aap~  380, 323

\bibitem[\protect\astroncite{{Labrosse} \& {Gouttebroze}}{2004}]{nl-lg04}
{Labrosse} N., {Gouttebroze} P., 2004,
  \apj~  617, 614

\bibitem[\protect\astroncite{{Labrosse} et~al.}{2006}]{nl-iau06}
{Labrosse} N., {Vial} J.~C., {Gouttebroze} P., 2006,
  Solar Active Regions and 3D Magnetic Structure, 26th IAU GA,
  JD 3, 47, 3

\end{thebibliography}
\end{document}